\documentclass{aa}

\usepackage{graphicx}
\usepackage{txfonts}
\usepackage[rflt]{floatflt}
\usepackage{natbib}        
\bibpunct{(}{)}{;}{a}{}{,} 

\newcommand{\base}{\ensuremath{\mathrm{B{}}}}
\newcommand{\wave}{\ensuremath{\lambda{}}}
\newcommand{\wavezero}{\ensuremath{\lambda_0{}}}
\newcommand{\deltawave}{\ensuremath{\Delta\lambda{}}}
\newcommand{\vis}{\ensuremath{\mu{}}}
\newcommand{\visSquare}{\ensuremath{\mu^2}}
\newcommand{\tfSquare}{\ensuremath{\mu^2_0{}}}
\newcommand{\blambda}{\ensuremath{\base{}_\wave{}}}

\newcommand{\blambdazero}{\ensuremath{\base^{_0}_\wave{}}}
\newcommand{\dUD}{\ensuremath{\small{\varnothing{}}}}
\newcommand{\airy}{\ensuremath{\mathcal{A}}}
\newcommand{\airySquare}{\ensuremath{\airy^2}}
\newcommand{\vSgr}{V3879~Sgr}
\newcommand{\nuLib}{$\nu$~Lib}

\begin{document}
\title{First result with AMBER+FINITO on the VLTI:\\ The high-precision angular diameter of V3879 Sgr\thanks{Based on observations collected during technical time at the European Southern Observatory, Paranal, Chile.}}
\titlerunning{First result with AMBER+FINITO on the VLTI}
\author{
  J.-B.~Le~Bouquin\inst{1}
  \and B.~Bauvir\inst{1,2}
  \and P.~Haguenauer\inst{1}
  \and M.~Sch\"oller\inst{1}
  \and F.~Rantakyr\"o\inst{1}
  \and S.~Menardi\inst{2}
}
\institute{
  European Southern Observatory, Casilla 19001, Santiago 19, Chile
  \and
  European Southern Observatory, Karl-Schwarzschild-Str. 2, 85748 Garching, Germany
}
\offprints{J.B.~Le~Bouquin\\
  \email{jlebouqu@eso.org}}
\date{Received; Accepted}
\abstract {}
{Our goal is to demonstrate the potential of the interferometric AMBER instrument linked with the Very Large Telescope Interferometer (VLTI) fringe-tracking facility FINITO to derive high-precision stellar diameters.}
{We use commissioning data obtained on the bright single star V3879~Sgr. Locking the interferometric fringes with FINITO allows us to record very low contrast fringes on the AMBER camera. By fitting the amplitude of these fringes, we measure the diameter of the target in three directions simultaneously with an accuracy of 25 micro-arcseconds.}
{We showed that V3879~Sgr has a round photosphere down to a sub-percent level. We quickly reached this level of accuracy because the technique used is independent from absolute calibration (at least for baselines that fully span the visibility null). We briefly discuss the potential biases found at this level of precision.}
{The proposed AMBER+FINITO instrumental setup opens several perspectives for the VLTI in the field of stellar astrophysics, like measuring with high accuracy the oblateness of fast rotating stars or detecting atmospheric starspots.}
\maketitle

\section{Introduction}

In optical, long-baseline interferometry, the most popular observable is the contrast (also called the visibility) of the interferometric fringes that appear when superposing the beams coming from several distant telescopes. The major limitation is the random optical delay introduced by the atmospheric turbulence, which make these fringes jitter around on the detector by a quantity larger than the fringe spacing. Practically, fringe blurring is avoided, either by reducing the integration time to a few milliseconds, which prevents observations of faint targets; or by using a dedicated fringe-tracking facility, the purpose of which is to stabilize fringes by measuring and correcting the optical delay in real-time. Fringe-tracking is generally used to observe fainter objects or to increase the spectral resolution, but this is not the only application. Another important gain is the possibility of recording very low-contrast fringes that that rise above the noise level after integration times longer than a few seconds.

Such small fringes are produced by astronomical objects spatially resolved by the interferometer. More precisely when observing a single star modeled by a Uniform Disk of diameter \dUD{}, the fringe visibility \vis{} drops according to the following law:
\begin{equation}
  \label{eq:0}
  \visSquare{} = \tfSquare(\wave,t) \;.\;\airySquare(\dUD{}\,\blambda{})
\end{equation}
where\begin{description}
\item[\blambda{}] is the spatial frequency of the observation, given by the ratio of the distance between the telescopes \base{} projected on the sky (called the baseline) and the observing wavelength \wave{}.
\item[\airy{}] is the Airy function (the Fourier transform of a disk of unitary size). The function \airySquare{} actually goes to zero (fringes disappear) for a disk of diameter $\dUD{}\approx{}1.22/\blambda{}$. After this first null the fringes reappear, but at lower visibility (second lobe).
\item[\tfSquare{}] is the interferometric response of the instrumental chain, also called the transfer-function. It changes with the instrumental setup and the atmospheric conditions. Therefore, it has to be frequently calibrated by observing unresolved stars, or stars with known diameters. Statistical errors and fluctuations of this transfer-function generally dominate the error budget when trying to estimate the diameter with high accuracy.
\end{description}
Because of the structure of Eq.~\ref{eq:0}, collecting visibility points in the vicinity of the first null of the \airySquare{} function makes the diameter estimation much less sensitive to the transfer-function uncertainty. For instance, measuring the exact spatial frequency at which the fringes disappear gives an estimation of \dUD{}, formally independent from \tfSquare{}, since one can use the relation: $\dUD{}\approx{}1.22/\blambda{}$.

In this paper, we demonstrate the feasibility of this technique at the Very Large Telescope Interferometer \citep[VLTI, see][]{Scholler-2006jul} by using an adequate setup of the fringe-tracker FINITO \citep{Gai-2004oct} and of the scientific instrument AMBER \citep{Petrov-2007mar}. Section~\ref{sec:observationsAndDataReduction} describes the instrumental setup, the observations, and the data reduction. Section~\ref{sec:resultsAndDiscussion} details the results and discusses the obtained accuracy. The paper ends with brief conclusions and perspectives.

\begin{figure} 
  \begin{flushright}  \includegraphics[scale=1]{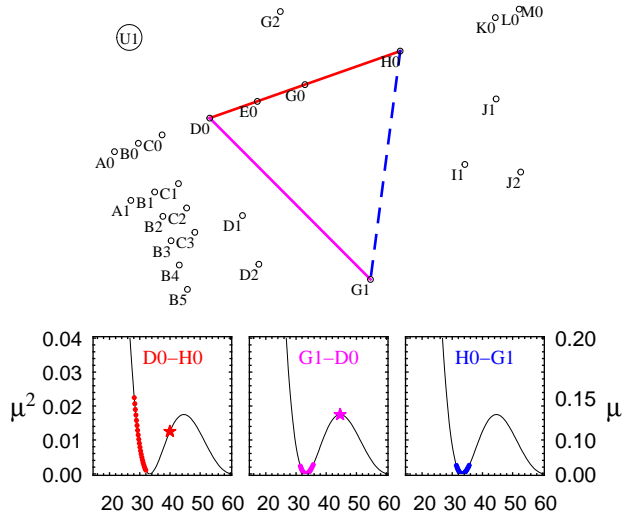} \end{flushright}
  \centering \caption{\label{fig:technique} VLTI array configuration during the observation (top) and corresponding simulated visibility-curves (solid line, bottom), assuming the uniform disk diameter of \vSgr{} ($7.56$~mas). FINITO tracks the fringes along D0-H0 and G1-D0 in the H-band (star symbols), while AMBER records data of the three baselines across the K-band (circle symbols). The horizontal axes are the spatial frequencies \blambda{}, marked in meter per micron, the vertical axes are the squared and linear visibilities.}
\end{figure}

\section{Observations and data reduction}
\label{sec:observationsAndDataReduction}

\subsection{Fringe tracking at VLTI}

A description of the VLTI fringe-tracking facility can be found in \citet{Gai-2004oct}. We recall here only its principle and main characteristics. The FINITO instrument records two baselines of a telescope triangle. Wide H-band interferometric fringes are formed by temporally modulating the optical paths before the beam combination. Output signals are processed in real-time with a modified ABCD algorithm. The measured fringe phases are used to reject fringe motion for frequencies below 20Hz. When the loop is closed, atmospheric perturbations are attenuated down to $0.15$\,$\mu$m RMS (performance routinely achieved on the 1.8m Auxiliary Telescopes but not yet on the larger 8.2m Unit Telescopes).

\subsection{Observations}

We obtained data at the VLTI during the night of 2007 May 1, with the relocatable $1.8$\,m Auxiliary Telescopes AT2, AT3, and AT4 placed respectively at stations G1, D0, and H0. Ground baseline length is 64m for baseline D0-H0, and 71m for baselines G1-D0 and G1-H0. AMBER was configured in medium resolution mode (R$\,\sim\,$1500) with a spectral window $1.95-2.25$\,$\mu$m. During the night, we recorded 22 files (or exposures) on the bright semi-regular pulsating star \vSgr{}. Hour angles of observations range from $-01$:$20$ to $01$:$15$. Each file is composed of 70 frames of 1s integration time each (instead of the 50ms integration time generally used when the fringes are not locked by FINITO). At the beginning of the observation, we also recorded 6 files with the same instrumental setup on the calibrator star \nuLib{}. The stellar parameters of \vSgr{} and \nuLib{} are summarized in Table~\ref{tab:stars}.
\begin{table}
  \centering
  \caption{Target parameters.}
  \begin{tabular}[c]{cccccccc}\hline\hline
    Target  & HD     & K    & Spectral & Published \dUD{} in K\\
    Name    & Number & (mag)& Type     & (mas)   \vspace{0.03cm}\\ \hline \vspace{-0.30cm}\\
    \vSgr{} & 172816 & -0.5 & M4III    & $8.09\pm{}0.27$ $^{1}$\\
    \nuLib  & 133774 & 1.61 & K5III    & $2.76\pm{}0.03$ $^{2}$\\ \hline
  \end{tabular}
  \flushleft
  \footnotesize{Spectral types and K-band magnitudes have been extracted from the Simbad database.
    Diameters have been extracted from:} \\
  \footnotesize{$1$: Lunar Occultation method \citep{Richichi-1998oct}} \\
  \footnotesize{$2$: Indirect method, reported in CHARM2 \citep{Richichi-2005feb}}
  \label{tab:stars}
\end{table}

Due to the combination of the \vSgr{} angular size, the baseline lengths and the FINITO working wavelength, we were able to lock the fringes near the second visibility-lobe maximum on baselines D0-H0 and G1-D0 (see Fig.~\ref{fig:technique}). Simultaneously, AMBER was recording data around the first visibility null with baselines H0-G1 and G1-D0, and at the very bottom of the first lobe with D0-H0 (which had a smaller projected baseline because of the star position in the sky). For a given baseline, AMBER records not a single point, but a range of \blambda{} because it spectrally disperses the light across the K-band. As expected, fringe contrasts estimated from the FINITO real-time display were in the range $10-15$\%, depending on the baseline and the hour angle. During the observations, FINITO provided a locking ratio systematically larger than 80\% (meaning that in each AMBER file, we recorded about 55 of the 70 frames with loop closed). However, we noticed that the FINITO performance degrades when the fringe visibility goes below $10$\%. We were not able to close the loop with visibilities smaller than $5$\%.

\subsection{Data reduction}
\label{sec:DataReduction}
\paragraph{Step 1:} We used the standard AMBER pipeline \texttt{amdlib} to convert the raw-data frames into uncalibrated square visibilities, following the reduction procedure described in detail by \citet{Tatulli-2007mar-All}. We kept the 50\% best frames sorted by signal-to-noise ratio on the fringe contrast. This efficiently discarded the frames recorded while the FINITO loop was open.

\begin{figure} 
  \centering 
  \includegraphics[scale=1]{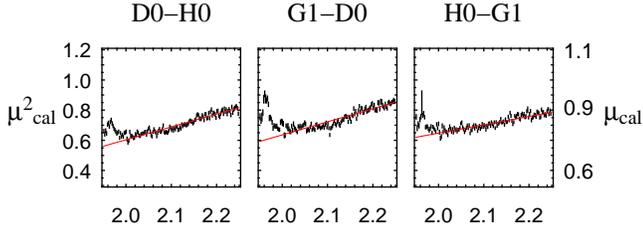}
  \caption{\label{fig:calibration} Transfer-function (observed visibilities calibrated from the stellar diameter of Table~\ref{tab:stars}) measured by AMBER on the reference star \nuLib{} in function of the wavelength in microns. The solid line is a linear fit of Eq.~\ref{eq:1} in the range 2.0-2.26\,$\mu$m.}
\end{figure}

\paragraph{Step 2:} To compute an estimation of the transfer-function, we fit the data obtained on the calibrator \nuLib{} by a function of the form:
\begin{equation}
  \label{eq:1}
  \visSquare_{cal} = \alpha{}\;( 1 + \beta{}\,\frac{\wave{}-\wavezero{}}{\deltawave})
\end{equation}
where \wavezero{} and \deltawave{} are the central and the full range of observed wavelengths; and $\visSquare_{cal}$ are the observed squared visibilities on \nuLib{}, calibrated from the stellar diameter of Table~\ref{tab:stars}. Data and associated fit are displayed in Fig.~\ref{fig:calibration}. We decided to restrict our analysis to the range 2.0-2.26\,$\mu$m in order to avoid the feature around 1.97\,$\mu$m. Discussing this feature is clearly out the scope of this paper, and more importantly this does not affect our results. We then removed the chromatic slope of the transfer-function from the \vSgr{} visibilities:
\begin{equation}
  \label{eq:2}
  \visSquare{} = \visSquare_{raw}\;/\;(1+\beta{}\,\frac{\wave{}-\wavezero{}}{\deltawave})
\end{equation}
where $\visSquare_{raw}$ are the observed squared visibilities on \vSgr{}. Note that we do not correct from the absolute value of the transfer-function, but only from its slope versus the wavelength. Therefore, our calibrated \vSgr{} visibilities are not \emph{equal}, but \emph{proportional} to the real visibilities, and this multiplicative factor should be independent from \wave{}.

\begin{figure} 
  \begin{flushright} \includegraphics[scale=1]{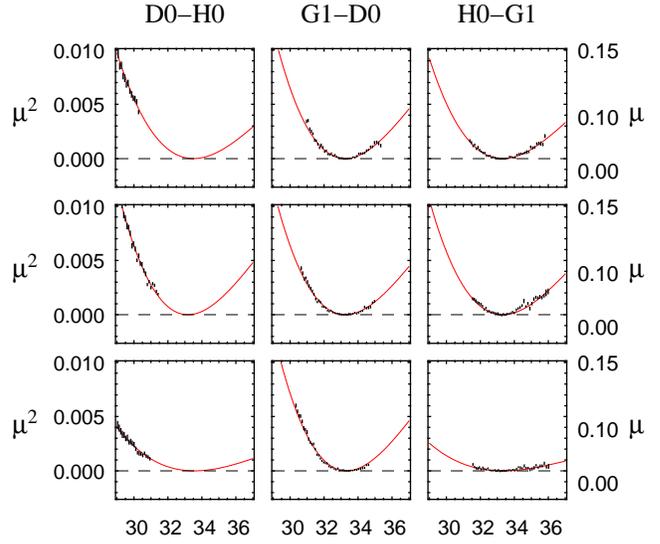} \end{flushright}
  \centering \caption{\label{fig:fit} Selected squared visibility curves. The data and the fit are shown for the three baselines and for three non-consecutive acquisitions (UT time 07:20, 09:45, and 10:03, from top to bottom). The horizontal axes are the spatial frequencies $\blambda$ marked in meter per micron, the vertical axes are the squared and linear visibilities (left and right scales respectively). The tracking performances on baseline D0-H0 degraded strongly at the very end of the night, explaining the visibility losses seen on D0-H0 and H0-G1 on the bottom line.}
\end{figure}

\paragraph{Step 3:} To extract uniform disk diameters, we fit independently each data set $\visSquare(\blambda)$ by Eq.~\ref{eq:0}, considering both the transfer-function \tfSquare{} and the diameter \dUD{} as free parameters. Because of the consecutive lobes of Eq.~\ref{eq:0}, the fit has several minima. To avoid being trapped by an obviously wrong minimum, we used as starting point the already published diameter $\dUD{}=8.09$\,mas and an ideal transfer-function $\tfSquare=1$.
During the fit process, we assumed that these two parameters do not vary with \wave{} (the star has a constant diameter and the slope of the transfer-function versus the wavelength has been corrected). Our fit also assumed the square visibility effectively to go down to zero between the first and the second lobe. This could be wrong if the target shape departs from central-symmetry. However, by looking at the data displayed in Fig.~\ref{fig:fit}, it is clear that the recorded minima on baselines H0-G1 and G1-D0 are fully compatible with zero ($\mathrm{min}(\visSquare{})=0\pm{}0.0002$), and therefore this assumption has no effect on the results. 

\begin{figure*} 
   \hspace{0.2cm}\includegraphics[scale=1]{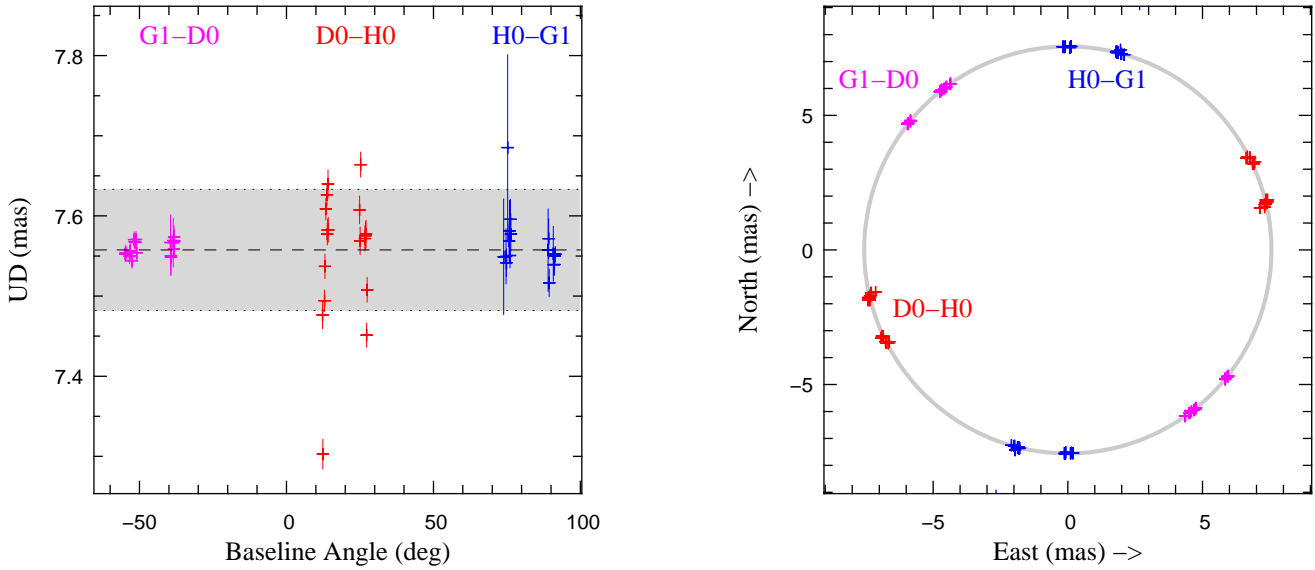}
  \centering \caption{\label{fig:diameters} Measured Uniform Disk diameter for \vSgr{}. Error bars do not take into account the potential systematic errors, but only the statistics coming from the squared visibility errors. In the right panel, the error bars are typically smaller than the symbol size. The shaded regions show the $\pm1$\% limits around the averaged size.}
\end{figure*}

Finally, statistical error-bars on the visibilities are propagated through the fit procedure to estimate the resulting uncertainties on the fit parameters $\dUD{}$ and $\tfSquare{}$. The formula assumed that 1) visibility errors are realistic and not cross-correlated, and 2) there is no other source of visibility error. The latter assumption is discussed in Sec.~\ref{sec:diameterAccuracy}. A sample of selected fits for different hour angles is displayed in Fig.~\ref{fig:fit}, and the computed diameters are displayed in Fig.~\ref{fig:diameters}. 

\section{Results and discussion}
\label{sec:resultsAndDiscussion}

Our results show that \vSgr{} has a round photosphere to sub-percent accuracy. Its diameter is about $7.56\pm{}0.025$\,mas, and this value is consistent in all measured directions. Because of its location in the sky, \vSgr{} has been previously observed several times with lunar occultation. We note that our value is about ten times more precise and within $2\sigma{}$ from the $8.09\pm{}0.27$\,mas reported by \citet{Richichi-1998oct}. In this work, \vSgr{} is reported as a visual variable ($\Delta=0.5^{\rm m}$ over a period of 50~days). Pulsations or a patchy atmosphere could explain the potential diameter discrepancy. Yet, systematic errors not taken into account in our computation and/or in the published values might also explain the discrepancy.

\subsection{Diameter accuracy}
\label{sec:diameterAccuracy}
On baseline G1-D0 the dispersion between the consecutive exposures is $\pm{}0.4$\% and is fully compatible with error bars computed by the fit process. On baseline H0-G1 the dispersion is $\pm{}1$\%, but is also fully compatible with the fit error bars. On the contrary, on baseline D0-H0 the dispersion is about $\pm{}1.5$\%, and is significantly larger than the fit error bars. Since FINITO was working on baselines D0-H0 and G1-D0, we can not blame the fringe-tracking loop or a complicated FINITO artifact in the AMBER data-reduction software.

\begin{figure} 
  \centering
  \vspace{0.25cm}
  \hspace{0.25cm} \includegraphics[scale=1]{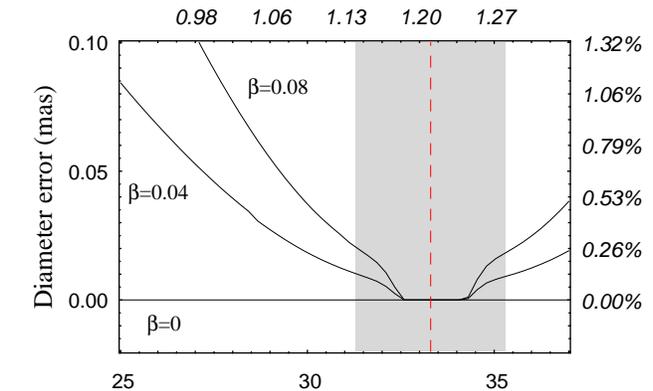}
  \centering \caption{\label{fig:sensitivity}
    Simulated bias on the diameter in function of the central spatial frequency (marked in meters per microns), and for various $\beta{}$ parameters. The gray zone are observations spanning the first visibility null (the visibility null is at $33.5\mathrm{\mu{}m/m}$ and the spatial frequency range is ${}5\mathrm{\mu{}m/m}$). Additional italic scales are dimensionless quantities: $\dUD{}.\blambda{}$ for the horizontal axis, and $\Delta\dUD{}/\dUD{}$ in percent for the vertical axis.}
\end{figure}

To test the influence of the calibration of the transfer-function slope on the results (Step~2 of Sect.~\ref{sec:DataReduction}), we simulated observations with an artificial visibility slope $\beta{}$ over the spatial frequency range. Fake data are given by the formula:
\begin{equation}
  \label{eq:3}
  \visSquare_{simu} =  \airySquare(\dUD{}\,\blambda)\;.\;(1 + \beta{}\,\frac{\blambda{}-{\blambdazero{}}}{\Delta\blambda{}})
\end{equation}
where ${\blambdazero}$ and $\Delta{}\blambda{}$ are the central and the full range of spatial frequencies of the simulation. We used a different definition than in Sec.~\ref{sec:DataReduction}, step~2 in order to define Eq.~\ref{eq:3} in a dimensionless manner. We chose the same target diameter, baseline length, spectral resolution and spectral window as for the real observations. Practically, this dimensionless computation remains valid as long as the ratio $\blambdazero{}/\Delta\blambda{}$ is constant ($\blambdazero{}/\Delta\blambda{}\approx 8.2$ in our data set). We then fit the simulated data with the same procedure as used in Step 3 of Sect.~\ref{sec:DataReduction}. The estimated diameter is displayed in Fig.~\ref{fig:sensitivity}, in function of the central spatial frequency $\blambdazero{}$ and for different values of $\beta{}$.

Clearly, a remaining instrumental visibility slope has little impact on the estimated diameter for observations spanning the visibility null. The latter acts as a strong ``locking point'' in the fit process. On the other side, when observing at $\blambda\sim{}27\mathrm{\mu{}m/m}$, as with the D0-H0 baseline, a change of $0.08$ in the slope of the transfer-function leads to an error of $0.1$mas (i.e. $1.3$\%) on the diameter.
Our favored explanation for the additional dispersion on baseline D0-H0 is a slight fluctuation of the transfer-function slope between the exposures (probably due to atmospheric seeing and coherence time). As a matter of fact, this also proves the robustness of the presented method on the two other baselines.

\subsection{Potential biases}
\label{sec:potentialBiases}
In our work, we see two specific potential biases, both related to the calibration of the spatial frequencies $\blambda{}$: the precision on the baseline length itself, and the precision on the AMBER spectral calibration. Note that these issues are generally ignored in optical interferometric data analysis since they are largely dominated by the transfer-function instabilities.

Concerning the baseline length, at VLTI, the three-dimensional vector between the two primary mirrors of a telescope pair is known with an accuracy of about one millimeter (conservative estimation). This corresponds to $\sim0.015$\% of the smallest baseline of our observations (60m). This value is significantly smaller than our computed statistical errors and can be completely ignored.
Another important issue related to the baseline length is the pupil transfer quality. If a telescope beam is vignetted on one side, the baseline is slightly shifted on the other side, since the effectively used section of the primary mirror is no longer centered around the mirror center (reference for the baseline model). In AMBER, the pupil is normally never vignetted more than $20$\%, which corresponds to a shift of a few tenths of the mirror size. If we assume a primary-mirror size of $1.8$\,m, and a shift of $10$\%, the relative error on the baseline is $\sim0.3$\%. This value is at the level of our precision.

Concerning the AMBER spectral calibration accuracy, the situation is less clear. This accuracy is not available at all in the literature. In a crude comparison of our AMBER spectra with atmospheric transmission curves in the K-band, we estimate the potential spectral error to be less than 0.02$\mu$m. This translates into a $1$\% error at 2\,$\mu$m (conservative estimation). This is significantly larger than our precision on the best baselines, meaning that our diameter estimation is most probably limited by this instrumental error source.

\section{Conclusions}

From the instrumental point of view, we have demonstrated that: \begin{itemize}
\item AMBER is able to record and process very low-contrast fringes when they are decently locked (down to about $1.5$\% contrast on a bright $\mathrm{K}=\-0.5^{\rm m}$ star, when integrating 60 frames of 1s each);
\item FINITO is able to lock fringes in the second visibility lobe (where fringes have less than $15$\% contrast) with a sufficiently good efficiency.
\end{itemize}
As an on-sky validation, we collected K-band interferometric fringes in the vicinity of the first visibility null of the bright star \vSgr{}. By fitting the visibility curves, we measured a diameter of $7.56\pm{}0.025$\,mas in three directions simultaneously. We reached such sub-percent accuracy without spending additional time on calibrators since the technique is independent from absolute calibration (at least for baselines that fully span the visibility null). We show that, at this level of precision, several systematic error sources have to be taken into account, for example the spectral calibration of the instrument and the pupil lateral position.

From the scientific point of view, this work opens several perspectives for the VLTI in the field of stellar astrophysics:\begin{itemize}
\item By using the AMBER+FINITO setup presented in this paper, one can measure a stellar diameter in three directions simultaneously with high accuracy. This can be used to constrain quickly the oblate photosphere of fast-rotating stars.
\item By using the same setup, one can precisely measure the value of the visibility minimum between the visibility lobes. Any departure from zero, even small, proves the presence of asymmetric features on (or above) the stellar surface, such as convective or magnetic starspots (in our \vSgr{} dataset, this minimum is compatible with zero, see Fig.~\ref{fig:fit}).
\item Finally, by associating fringe-tracking in the second lobe with baseline bootstrapping, one can record visibility points in the third visibility lobe. There, observables become very sensitive to the limb-darkening strength and its spectral dependency, which will allow thorough tests of stellar atmosphere models.
\end{itemize}

\begin{acknowledgements} 
  The authors want to warmly thank the complete VLTI team (Science Operations astronomers and engineers) for all the work accomplished on the FINITO and AMBER instruments. This work is based on observations made with the European Southern Observatory telescopes obtained from the ESO/ST-ECF Science Archive Facility. This research has made use of the Smithsonian/NASA Astrophysics Data System (ADS) and of the Centre de Donnees astronomiques de Strasbourg (CDS). All calculations and graphics were performed with the freeware \texttt{Yorick}\footnote{\texttt{http://yorick.sourceforge.net}}.
\end{acknowledgements}


\bibliographystyle{/Users/jlebouqu/Tex/AandA/bibtex/aa}  



\end{document}